\documentclass[preprint,superscriptaddress,nofootinbib]{revtex4-1}
\usepackage{hyperref}
\usepackage{bbm}
\usepackage{amsfonts}
\usepackage{mathrsfs}

\usepackage{amsmath,amssymb}

\usepackage{epsfig}
\usepackage{graphicx}               % Standard graphics package
\usepackage{url}
\usepackage{hyperref}
\usepackage{float}
\usepackage{pstricks}
\usepackage{color}
\usepackage{multirow}

\newcommand{\lsim}{\!\mathrel{\hbox{\rlap{\lower.55ex \hbox{$\sim$}} \kern-.34em \raise.4ex \hbox{$<$}}}}
\newcommand{\gsim}{\!\mathrel{\hbox{\rlap{\lower.55ex \hbox{$\sim$}} \kern-.34em \raise.4ex \hbox{$>$}}}}

\def\eg{{\it e.g.}}
\def\ie{{\it i.e.}}
\def\be{\begin{equation}}
\def\ee{\end{equation}}
\def\bea{\begin{eqnarray}}
\def\eea{\end{eqnarray}}
\def\bit{\begin{itemize}}
\def\eit{\end{itemize}}

\newcommand{\Eref}[1]{Eq.~(\ref{#1})}
\newcommand{\Fref}[1]{Fig.~\ref{#1}}
\newcommand{\Sref}[1]{Sec.~\ref{#1}}
\newcommand{\Tref}[1]{Tab.~\ref{#1}}
\newcommand{\Srefs}[2]{Secs.~\ref{#1} and~\ref{#2}}
\newcommand{\Erefs}[2]{Eqs.~(\ref{#1}) and~(\ref{#2})}
\newcommand{\Ref}[1]{Ref.~\cite{#1}}
\newcommand{\Refs}[1]{Refs.~\cite{#1}}
\newcommand{\vev}[1]{ \left\langle {#1} \right\rangle }
\newcommand{\abs}[1]{ \left| {#1} \right| }

\newcommand{\GeV}{\text{ GeV}}

\newcommand{\al}{\abs{\lambda}}

\def\ie{{\it i.e.}}
\def\eg{{\it e.g.}}

\begin{document}

\begin{flushright}
\text{\footnotesize FERMILAB-PUB-15-090-T}
\end{flushright}
\vskip 10 pt

\title{Is a Higgs Vacuum Instability Fatal for High-Scale Inflation?}
\author{John Kearney}
\affiliation{
Theoretical Physics Department \\
\vskip -8 pt
Fermi National Accelerator Laboratory, Batavia, IL 60510}
\author{Hojin Yoo}
\affiliation{Theory Group \\ \vskip -8 pt Lawrence Berkeley National Laboratory, Berkeley, CA 94709 USA}
\affiliation{Berkeley Center for Theoretical Physics \\ \vskip -8 pt University of California, Berkeley, CA 94709 USA}
\author{Kathryn M. Zurek}
\affiliation{Theory Group \\ \vskip -8 pt Lawrence Berkeley National Laboratory, Berkeley, CA 94709 USA}
\affiliation{Berkeley Center for Theoretical Physics \\ \vskip -8 pt University of California, Berkeley, CA 94709 USA}

\begin{abstract}

We study the inflationary evolution of a scalar field $h$ with an unstable potential for the case where the Hubble parameter $H$ during inflation is larger than the instability scale $\Lambda_I$ of the potential.  Quantum fluctuations in the field of size $\delta h \sim \frac{H}{2 \pi}$ imply that the unstable part of the potential is sampled during inflation.  We investigate the evolution of these fluctuations to the unstable regime, and in particular whether they generate cosmological defects or even terminate inflation.
We apply the results of a toy scalar model to the case of the Standard Model (SM) Higgs boson, whose quartic evolves to negative values at high scales, and extend previous analyses of Higgs dynamics during inflation utilizing statistical methods to a perturbative and fully gauge-invariant formulation.  We show that the dynamics are controlled by the renormalization group-improved quartic coupling $\lambda(\mu)$ evaluated at a scale $\mu = H$, such that Higgs fluctuations are enhanced by the instability if $H > \Lambda_I$.
Even if $H > \Lambda_I$, the instability in the SM Higgs potential does not end inflation; instead the universe slowly sloughs off crunching patches of space that never come to dominate the evolution.  As inflation proceeds past 50 $e$-folds, a significant proportion of patches exit inflation in the unstable vacuum, and as much as 1\% of the spacetime can rapidly evolve to a defect.
Depending on the nature of these defects, however, the resulting universe could still be compatible with ours.

\end{abstract}

\preprint{}

\maketitle

%%%%%%%%%%%%%%%%%%%%%%%%%%%%%%%%
\section{Introduction}
\label{sec:introduction}
%%%%%%%%%%%%%%%%%%%%%%%%%%%%%%%%

The discovery of the Higgs boson with mass 125 GeV and Standard Model (SM)-like couplings at the Large Hadron Collider \cite{Aad:2012tfa,Chatrchyan:2012ufa} has ushered in a new era in particle physics, with particular emphasis on studying the properties of the Higgs field.  One of the most important aspects of a SM Higgs boson with the observed mass is that its potential becomes unstable at high scales---the quartic coupling $\lambda$ in the potential $V(h) \approx \frac{\lambda(\mu)}{4} h^4$ (ignoring the mass-squared term, negligible at high energy) evolves through the renormalization group equations to negative values at scales $\mu > \Lambda_I$.  In the SM, $\Lambda_I$ is in the range $10^{9}$-$10^{16}$ GeV (within $2 \sigma$ uncertainties on the top and Higgs masses).

One implication of this instability is that our Universe (with the Higgs and SM only) is metastable, meaning that, while the electroweak vacuum is stable on time scales longer than the age of the Universe, it can ultimately decay through a Coleman-De Luccia instanton \cite{Coleman:1980aw} at late times.  The presence of this instability has been known a long time \cite{Sher:1988mj,Casas:1994qy,Altarelli:1994rb,Sher:1993mf}, and has been explored more recently in, \eg, \cite{Isidori:2001bm,Isidori:2007vm,Ellis:2009tp,EliasMiro:2011aa}.\footnote{Previously neglected effects that could potentially alter the results of the standard metastability calculation have also recently been explored in \cite{Branchina:2013jra,Branchina:2014usa,Branchina:2014rva,Lalak:2014qua,Burda:2015isa,Burda:2015yfa}.}
While an interesting observation, the metastability of the electroweak vacuum with a 125 GeV Higgs boson does not phenomenologically impact the existence of our Universe at the present time.

The instability in the Higgs potential may be more relevant, however, in influencing the evolution of the Universe during inflation.  This is because quantum fluctuations in the Higgs field during inflation, $\delta h \sim \frac{H}{2 \pi}$ (where $H$ is the Hubble parameter during inflation), can locally drive the Higgs vacuum expectation value (vev) to the unstable part of the potential---in particular, if the instability scale is relatively low, $\Lambda_I \lsim 10^{14} \GeV$, this can readily occur even for modest values of $H \gsim \Lambda_I$. Consequently, the existence of the instability could imply a constraint on $H$ or the form of the Higgs potential during inflation, and may be particularly relevant if a large tensor-to-scalar ratio $r \sim 0.1$, corresponding to $H \sim 10^{14} \GeV$, is observed in future cosmic microwave background (CMB) experiments.  This has been the subject of a number of papers \cite{Espinosa:2007qp,Choi:2012cp,Lebedev:2012sy,Kobakhidze:2013tn,Enqvist:2013kaa,Fairbairn:2014zia,Hook:2014uia,Herranen:2014cua,Kobakhidze:2014xda,Kamada:2014ufa,Torabian:2014nva,Enqvist:2014bua,Shkerin:2015exa}.

Because of the scale dependence of the Higgs potential, the nature of the Higgs as a field that breaks electroweak symmetry, and the fact that inflation creates causally disconnected regions of space that are free to evolve independently, a number of subtleties must be addressed in order to correctly study the Higgs field during inflation.  These complications impact the result substantially and, as a result, there is disagreement between various results in the literature.

First, one must understand what transition from the electroweak to unstable vacuum during inflation means physically for the existence of our Universe.  During inflation, spacetime that eventually becomes part of our Universe is continually passing out of causal contact---inflation is based on the idea that a single homogeneous patch evolves into $e^{3{\cal N}}$ distinct Hubble volumes, where ${\cal N} \gsim 50$ is the number of $e$-folds of inflation required to obtain a Universe flat and homogeneous like ours.  Thus, even if a single Hubble patch (of size $H^{-1}$ in physical coordinates) transitions to the unstable vacuum during inflation or ultimately crunches due to the negative energy density in the Higgs field, the background is still dominated by the inflaton and causally disconnected patches still undergo exponential expansion.  So, one expects the spacetime to continue to inflate globally, and any resulting defect to be inflated away.
For these reasons it is unclear, in contrast to the assumption of \cite{Fairbairn:2014zia}, why a single local fluctuation of the Higgs vev to the unstable regime during inflation would be fatal for our Universe.
Only if a significant fraction of the $\sim e^{3 {\cal N}}$ Hubble volumes crunch (in particular, near the end of inflation when they are not diluted by further expansion of space) does one expect the resulting large inhomogeneities to potentially be inconsistent with the small perturbations observed in our Universe.
Meanwhile, if patches exhibiting a Higgs vev $\gsim \Lambda_I$ are present at the end of inflation and not stabilized by, \eg, reheating, these patches could destroy the patches of electroweak vacuum as the various patches come back into causal contact.
The proportion of Hubble volumes that transition and when they transition is thus important, and one must appropriately evolve the Higgs field during inflation to evaluate the existence of a Universe like ours.

Many approaches have employed the Hawking-Moss instanton calculation \cite{Hawking:1981fz} to determine the probability that the Higgs transitions to the top of the potential barrier (from where it is assumed to subsequently evolve towards the true vacuum) during an $e$-fold of inflation, $P \sim \exp(-\frac{8 \pi^2 \Delta V}{3 H^4})$ \cite{Kobakhidze:2013tn,Herranen:2014cua,Kobakhidze:2014xda}.
Here, $\Delta V = V(h=\Lambda_{\rm max}) - V(0)$, where $V(h = \Lambda_{\rm max})$ is the maximum value of the Higgs effective potential---in the SM, $h = \Lambda_{\rm max}$ occurs just before the quartic becomes negative.
This approach would in principle be suitable to calculate, \eg, the proportion of patches at the top of the barrier during a given $e$-fold.
However, as the preceding discussion indicates, in order to understand the evolution of the space as a whole we are interested in the full distribution of the Higgs vev values during and at the end of inflation.  Furthermore, for large $H^4 \gg \Delta V$ (as for the SM Higgs potential with $H \gsim \Lambda_I$), the exponent goes to zero and the unknown prefactor becomes important.
As a result, the HM calculation is insufficient to fully study the dynamics of the Higgs field during inflation.

An alternative approach, valid when $H^4 \gsim \Delta V$, remedies these problems by allowing for a thermal diffusion of the field in its potential, with a temperature characterized by the de-Sitter temperature $T_{\rm dS} = \frac{H}{2\pi}$.  Such a treatment is encapsulated in the Fokker-Planck equation \cite{Espinosa:2007qp,Fairbairn:2014zia,Hook:2014uia}.  By treating a statistical ensemble of baby universes, a probability distribution $P(h,t)$ for the Higgs vev $h$ in a patch is derived as a function of the duration of inflation.
While it has been argued that the Fokker-Planck approach reproduces a Hawking-Moss-like distribution in the late-time equilibrium limit (see, \eg, \cite{Starobinsky:1994bd}), the Fokker-Planck equation incorporates dynamics not captured by the Hawking-Moss calculation.
As a result, the Fokker-Planck treatment is more suitable for studying the behavior of the Higgs field during inflation and the implications for our Universe.
While \Ref{Fairbairn:2014zia} restricted their focus to the case that the transition of a single Hubble patch in our past light cone to the unstable vacuum destroys the entire Universe,
\Ref{Espinosa:2007qp}, in contrast, assumed that any patches transitioning to the unstable regime during inflation rapidly but benignly crunch.  They thus focused on the proportion of electroweak patches that survive until the end of inflation $t_e$,
\be
\label{eq:survivalprobability}
P_\Lambda \equiv \int_{- \Lambda_{\rm max}}^{\Lambda_{\rm max}} d h \, P(h,t_e).
\ee
To calculate $P_\Lambda$, they solved the FP equation subject to the boundary condition $P(\abs{h} = \Lambda_{\rm max},t) = 0$ on the grounds that, for $\abs{h} \geq \Lambda_{\rm max}$, the unstable potential causes $\abs{h}$ to roll off to infinity and the patch to crunch.  However, the background energy density in patches with $\abs{h} \geq \Lambda_{\rm max}$ is generally still dominated by the inflaton, and rapid runoff of $\abs{h}$ and flattening of $P$ only occurs once the classical force due to the unstable potential dominates over the quantum fluctuations.  This happens when
\be
h \gsim h_{\rm classical} \equiv \left(\frac{3}{-2 \pi \lambda}\right)^{1/3} H \gg \Lambda_{\rm max}.
\ee
As such, imposing $P(\abs{h} = \Lambda_{\rm max},t) = 0$ artificially suppresses the probability to find $\abs{h} \sim \Lambda_{\rm max}$, and hence $P_{\Lambda}$.

\Ref{Hook:2014uia} corrected this unphysical boundary condition by solving the Fokker-Planck equation with boundary condition $P(\abs{h} = \Lambda_c,t) = 0$ for $\Lambda_c \geq h_{\rm classical}$, which captures the bulk of the distribution at $\abs{h} < h_{\rm classical}$ and so is suitable for calculating $P_{\Lambda}$ in the case $H > \Lambda_I$.\footnote{The $P_{\Lambda}$ calculated in \cite{Hook:2014uia} were independent of the exact choice of $\Lambda_c$ provided it was chosen to be above where the distribution rapidly flattens for $\abs{h} \geq \Lambda_c \geq h_{\rm classical}$, consistent with the observation therein that the boundary condition employed to solve the FP equation is unphysical.}
This change drastically increased the electroweak survival probability and consequently revealed that, provided patches in the unstable regime do crunch benignly, inflation can always last long enough to replace the lost patches.
So, the instability in the SM Higgs potential does not necessarily preclude the existence of our Universe in this case.

While laying to rest the question of transition probabilities during inflation, \Ref{Hook:2014uia} left some questions unanswered.  For instance, \Ref{Hook:2014uia} remained agnostic as to the exact implications of $P(h,t_e)$ for our Universe, considering the two limiting cases of ``benign crunching'' and that a single unstable patch in our past light-cone destroys all electroweak patches.
In addition, a number of technical points remained unclear.  For instance, there is the question of which potential $V(h)$ to use in solving the evolution equation.
For the Higgs boson, one is tempted to use the effective potential $V_{\rm eff}$ as computed in, \eg, \cite{Buttazzo:2013uya}, or the appropriate analog in de Sitter (dS) space as computed in, \eg, \cite{Herranen:2014cua}.  The value of $V_{\rm eff}(h)$, however, is gauge-dependent (except at stationary points) \cite{Nielsen:1975fs}.  Consequently, as recently emphasized in \Refs{Andreassen:2014gha,Andreassen:2014eha,DiLuzio:2014bua}, one must be careful in extracting physical quantities from the effective potential.
Furthermore, field values such as $\Lambda_{\rm max}$ are gauge-dependent, potentially presenting a problem when attempting to determine whether or not a patch has fluctuated to the unstable regime.
Note, though, that this caveat does not necessarily imply that the survival probability defined in \Eref{eq:survivalprobability} is unphysical, as it is based on relative field values---the field is simply assumed to evolve to the electroweak (unstable) regime if $\frac{\abs{h}}{\Lambda_{\rm max}} < \, (>) \, 1$ at $t_e$.

The purpose of this paper is to address these outstanding issues for the case of the Higgs boson, including the viability of our Universe given the SM Higgs potential instability and a gauge-invariant evolution of the field---we will find several conceptual improvements over \cite{Hook:2014uia}, though numerically our results are unchanged.
To do this, we first set aside the Fokker-Planck approach, and consider the evolution and growth during inflation of the two-point correlation function of scalar field fluctuations for a toy model with $V(h) = \frac{\lambda}{4} h^4$, assuming $\lambda < 0$ such that the potential is unstable.  This will elucidate both how to extend such a model to the case of the physical Higgs boson in an appropriately gauge-invariant and physical way, and how to capture the effects of the full SM in such a model.  Consequently, it will assist us in correctly interpreting the results of the Fokker-Planck calculation for the SM Higgs.

Specifically, in \Sref{sec:hartreefock}, we do a mode decomposition of $h$, integrate out the sub-horizon modes and compute the evolution of the vev fluctuations, assuming a Gaussian distribution.  In doing so, we gain an important physical insight: the variance of the distribution becomes infinitely broad after a finite number of $e$-folds of inflation.  This indicates that, during each subsequent $e$-fold, we expect the vev in a significant proportion of surviving patches to rapidly diverge, giving rise to a sizable number of crunching patches and defects.  If inflation were to successfully end after this point, the resulting Universe would likely exhibit large inhomogeneities, and consequently look rather unlike ours.
With this physical insight, in \Sref{sec:2pt} we recalculate the two-point correlation function for the fluctuations in perturbation theory.
Doing so reveals that a stochastic approach, such as that employed in \Sref{sec:hartreefock}, captures the leading, gauge-invariant contributions to the correlator provided certain identifications are made.
In particular, we see that the quartic is a function of scale $\mu$, $\lambda(\mu)$, which must be fixed in the calculation.
The perturbative calculation shows that the appropriate coupling to use to study the evolution of the Higgs field is the renormalization group (RG)-improved Higgs quartic coupling evaluated at the Hubble scale during inflation, $\lambda(H)$.  This is gauge-invariant, and hence physical.  Moreover, it encapsulates the sub-horizon effects of the SM gauge bosons and fermions.
The perturbative calculation also reveals how to treat the additional degrees of freedom in the full SM Higgs doublet.
Armed with this enhanced understanding, we return in \Sref{sec:FP} to the Fokker-Planck equation, and use these results to interpret the resulting probability distribution function for the SM Higgs boson and hence our Universe.  Finally, we conclude.

%%%%%%%%%%%%%%%%%%%%%%%%%%%%%%%
\section{Toy Model: $\lambda h^4$ Field Evolution in the Gaussian Approximation}
\label{sec:hartreefock}
%%%%%%%%%%%%%%%%%%%%%%%%%%%%%%%

We begin by calculating the evolution of a scalar field in dS space employing a toy model frequently used in the literature and outlined in \cite{Starobinsky:1994bd}.   This model illustrates many of the important features, and serves as a check on the results, of the full SM Higgs case analyzed in \Srefs{sec:2pt}{sec:FP}.   It consists of a quartically-coupled real scalar,
\be
V(h) = \frac{\lambda}{4} h^4
\ee
where $\lambda$ is taken to be constant.  This simple model will turn out to be a good approximation for the Higgs field during the early stages of inflation, provided $\lambda$ is chosen appropriately.  In the case of the Higgs, the value of the coupling $\lambda(\mu)$ depends on the relevant energy scale---we will see in the next section that an appropriate choice is $\mu = H$, and here we implicitly consider $\lambda < 0$ such that the above potential is unstable as for the Higgs field during a period of inflation with $H > \Lambda_I$.  In addition, we assume the scalar $h$ is minimally-coupled and that its potential does not receive large corrections due to the inflaton energy density.  Non-minimal curvature coupling \cite{Herranen:2014cua,Enqvist:2014bua}, coupling to the inflaton \cite{Lebedev:2012sy} or higher-dimension operators \cite{Hook:2014uia} can serve to stabilize or destabilize the potential during inflation.  Within the context of this simplified model we show that the correlation function for the scalar field fluctuations, $\vev{\delta h^2(t)}$, diverges in finite time, and we discuss the implications of this divergence for our Universe.

The equation of motion for a canonically-normalized scalar field $h$ in a dS background is given by
\be
\label{eq:higgseom}
\ddot{h} + 3 H \dot{h} - \left(\frac{\vec{\nabla}}{a}\right)^2 h + V'(h) = 0.
\ee
We decompose the scalar field in terms of a homogeneous background value $\bar{h}(t)$ and local fluctuations $\delta h(x,t)$.  We will assume $\bar{h}(0) = 0$, $\bar{h}(t) = 0$ throughout inflation; taking non-zero values will only lead to faster divergence.  In this case, \Eref{eq:higgseom} is the equation of motion for the fluctuations of the Higgs field, which can be decomposed into mode functions
\be
\label{eq:higgsmodes}
\delta h(x,t) = \int \frac{d^3 k}{(2\pi)^3} a_{\vec{k}} \delta h_k(t) e^{i \vec{k} \cdot \vec{x}} + \text{h.c.},
\ee
where the creation and annihilation operators $a_{\vec{k}}, {a_{\vec{k}}}^\dagger$ satisfy the usual communtation relations.

We now consider the evolution of the fluctuations in the context of the Hartree-Fock (HF) or Gaussian approximation, where we can write all higher-point correlators in terms of $\vev{\delta h^2(t)}$.  As we discuss in \Sref{sec:FP}, this is a good approximation before fluctuations become large and self-interactions become relevant.
Using the Gaussian approximation we can linearize \Eref{eq:higgseom}, including the interactions, and then inserting \Eref{eq:higgsmodes} into \Eref{eq:higgseom} gives the mode equation
\be
\label{eq:modeeom}
\ddot{\delta h}_k(t) + 3 H \dot{\delta h}_k(t) + \left\{\left(\frac{k}{a}\right)^2 + 3 \lambda \vev{\delta h^2(t)}\right\} \delta h_k(t) = 0,
\ee
where
\be
\label{eq:twopointdef}
\vev{\delta h^2(t)} = \int_{k = 1/L}^{k = \epsilon a H} \frac{d^3 k}{(2\pi)^3} \abs{\delta h_k(t)}^2
\ee
is the two-point correlation function for the value of the scalar field in a Hubble patch, obtained by integrating over all superhorizon modes with $k \leq \epsilon a H$.  $\epsilon$ is an $\mathcal{O}(1)$ constant chosen to distinguish between sub- and superhorizon modes, though our results will ultimately be independent of $\epsilon$. We will take $t_k$ to be the time that the physical wavelength of the mode exceeds the horizon size and the mode freezes out, given by $k = \epsilon a(t_k) H$.

In writing \Eref{eq:modeeom} with the integral of \Eref{eq:twopointdef} taken over superhorizon modes only, we have neglected subhorizon mode correlations.  These terms can be cancelled using local counterterms in order to derive an equation describing the evolution of superhorizon modes, and as such the dominant effects of subhorizon modes can be reabsorbed into renormalization of the coupling $\lambda$---we return to this point in \Sref{sec:2pt}.
In addition, note that \Eref{eq:twopointdef} requires an infrared (IR) cutoff, corresponding to the fact that we are studying fluctuations relative to a homogeneous background value and so only consider modes that were subhorizon at the onset of inflation.  We choose a co-moving box of length $L$ whose size is simply given by the region of space over which the initial condition $\bar{h}(0) = 0$ is a good approximation, corresponding to an IR cutoff $k \ge a_0 H$ where $a_0$ is the scale factor as the onset of inflation.\footnote{The modes with $k < a_0 H$ effectively determine $\bar{h}(0)$ within this box.}  The IR cutoff corresponds to the longest observable scale (or ``resolution'') for observing mode fluctuation relative to a homogeneous background value, which is limited to a causally-connected region at the beginning of inflation \cite{Lyth:2007jh}.

Now, consider the evolution of $\vev{\delta h^2(t)}$ assuming that
\be
\al \vev{\delta h^2(t)} \ll H^2.
\ee
In this case, modes are effectively massless, yielding the usual result in dS space
\be
\label{eq:modesolution}
\delta h_k(t) = \frac{H}{\sqrt{2 k^3}}\left(1-i\frac{k}{aH}\right) e^{i\frac{k}{aH}}.
\ee
For the slowly-varying superhorizon modes with $1/L \leq k \leq \epsilon a H$, we can employ the slow-roll approximation and neglect the gradient term such that
\be
\label{eq:shmodeeom}
3 H \dot{\delta h}_k(t) + 3 \lambda \vev{\delta h^2(t)} \delta h_k(t) = 0.
\ee
The evolution equation for $\vev{\delta h^2(t)}$ can be found by multiplying by $\delta h^\ast_k(t)$ and integrating over superhorizon modes.  The derivative term is simplified using
\be
\int \frac{d}{dt} \abs{\delta h_k(t)}^2 = \frac{d}{dt} \left(\int \abs{\delta h_k(t)}^2\right) - \frac{4 \pi k^2}{(2 \pi)^3} \abs{\delta h_k(t_k)}^2 \frac{d}{dt} (\epsilon a H) = \frac{d}{dt} \vev{\delta h^2(t)} - \frac{H^3}{4 \pi^2}.
\ee
We pick up a stochastic or Brownian noise term as a result of the time-dependence of mode horizon crossing.  This derivation is one method by which to obtain the well-known result that de Sitter space behaves thermally.
The equation governing the evolution of $\vev{\delta h^2(t)}$ is then
\be
\label{eq:hsqevolution}
\frac{d}{dt} \vev{\delta h^2(t)} = -\frac{2 \lambda}{H} \vev{\delta h^2(t)}^2 + \frac{H^3}{4 \pi^2}.
\ee
The solution to this equation is
\be
\label{eq:HFSoln}
\vev{\delta h^2(t)} = \frac{1}{\sqrt{- 2 \lambda}} \frac{H^2}{2 \pi} \tan\left(\sqrt{- 2 \lambda} \frac{{\cal N}}{2 \pi}\right)
\ee
where ${\cal N} = Ht$ is the number of $e$-folds of inflation.  While we have written the result as for $\lambda < 0$, it is equally valid for $\lambda > 0$ (with the tangent function replaced by a hyperbolic tangent).

There are several notable features of \Eref{eq:HFSoln}.  First, in the limit of $\lambda \rightarrow 0$, we obtain the familiar result for a massless field in dS space
\be
\label{eq:masslessDSresult}
\vev{\delta h^2(t)} = \frac{H^2 {\cal N}}{4 \pi^2}.
\ee
Second we observe that, for $\lambda > 0$, the interaction tends to reduce the size of the fluctuations and stabilize the scalar field---the distribution of field values approaches an equilibrium state, as described in \cite{Starobinsky:1994bd}.  The more interesting case is when $\lambda < 0$, as for the SM Higgs with $H > \Lambda_I$ such that $\lambda(H) < 0$.  In this case, we see that the superhorizon fluctuations grow even more rapidly than for a massless field, and in fact diverge after a finite number of $e$-folds,
\be
\label{eq:Nmax}
{\cal N}_{\rm max} = \frac{\pi^2}{\sqrt{- 2 \lambda}}.
\ee

What does this divergence mean physically?  As mentioned previously, $\vev{\delta h^2(t)}$ is the correlation function for local superhorizon mode fluctuations (``local'' meaning the field value is averaged over a Hubble-sized patch).  It is analogous to more familiar correlation functions such as $\vev{\delta \phi^2(t)}$, where $\phi$ is the inflaton and $\delta \phi$ represents the local quantum fluctuations around the homogeneous background value.  In the same way that the local fluctuations in the inflaton value give rise to local fluctuations in energy density, the fluctuations $\delta h(x,t)$ give rise to different values of the field value in different patches and hence different local energy densities.

If the field value in a particular patch fluctuates to a very large value such that $\abs{\lambda} \delta h^4 \gsim H^2 M_P^2$, the energy density in the field $\rho_h \approx V(\delta h) < 0$ may cancel the inflaton energy density $\rho_\phi \sim H^2 M_P^2$, producing a patch in which the local energy density is small or negative.
This backreaction causes the patch to stop inflating and crunch, giving rise to a defect such as a black hole.
More precisely, solving the Friedmann equations reveals that, once the field value in a patch exits the slow-roll regime, $\abs{\delta h} \gsim \sqrt{\frac{3}{-\lambda}}$, the field value diverges rapidly and the patch quickly evolves to a singularity, within $\sim 1$ $e$-fold.
In the Gaussian approximation, though, the typical size of a field fluctuation in a patch is of order $\sqrt{\vev{\delta h^2(t)}}$.  Consequently, such large fluctuations are extremely rare throughout most of inflation.
Moreover, the rare occurrence of backreacting and non-inflating patches does not disrupt inflation globally, and the resulting defects would be diluted by inflation, minimizing observational implications.

However, when ${\cal N}$ approaches ${\cal N}_{\rm max}$, large field value fluctuations are no longer rare; a significant fraction of the patches that eventually evolve into the observable Universe would develop instabilities.  Consequently, the resulting Universe would exhibit large inhomogeneities as a result of the defects produced---in the case of our Universe, large inhomogeneities would be inconsistent with the small curvature perturbations $\Delta_R^2 \approx 2 \times 10^{-9}$ observed by, \eg, WMAP \cite{Hinshaw:2012aka} and Planck \cite{Planck:2015xua}.  In addition, if the proportion of non-inflating patches becomes ${\cal O}(1)$, the analysis of \cite{Sekino:2010vc} suggests that the inflating regions could not percolate and undergo the necessary amount of inflation (inflating regions would fracture or ``crack'').  The inflating space as a whole becomes unstable, and inflation ends.  Thus if ${\cal N}_{\rm max} \lsim {\cal N}_o$, where ${\cal N}_o \approx 50-60$ is the number of $e$-folds needed to satisfy observational bounds on flatness and homogeneity, then a relatively homogeneous Universe such as ours would not be consistent with the existence of a scalar field such as $h$ exhibiting an instability in its potential.  We will return to this point in \Sref{sec:FP}.

Consequently, having ${\cal N}_o \leq {\cal N}_{\rm max}$ is necessary, but not sufficient, to guarantee the existence of our Universe.  After inflation ends, rapid reheating must occur to stabilize the potential and prevent collapse of the entire spacetime.  Finite temperature effects generate a positive mass-squared for $h$, $m_{eff}^2 \sim T_R^2$, where $T_R$ is the re-heat temperature.  As long as $m_{eff}^2 \gtrsim \lambda \langle \delta h \rangle^2$, the field is rapidly thermalized and driven to $\langle \delta h^2 \rangle = 0$.  This is easily satisfied, since the maximum re-heat temperature is $T_R^{\rm max} \sim \sqrt{H M_P}$, while $\langle \delta h^2 \rangle$ is typically of size $\frac{H^2{\cal N}}{(2 \pi)^2}$.

In deriving these results, we have employed several approximations.
First, we have assumed $\al \vev{\delta h^2(t)} \ll H^2$, such that the fluctuations are effectively massless and the evolution of the superhorizon modes can be considered in the slow-roll approximation.  If $\lambda < 0$, modes become tachyonic once $\al \vev{h^2(t)} \gsim H^2$, leading to their rapid growth.  This coupled with the rapid (\ie, not slow-roll) evolution of superhorizon modes in this regime accelerates the divergence of field fluctuations, making the above estimate of ${\cal N}_{\rm max}$ an upper bound.  However, the accelerated growth of $\vev{\delta h^2(t)}$ near ${\cal N}_{\rm max}$ means that this assumption is not violated significantly before $\vev{\delta h^2(t)}$ diverges, such that ${\cal N}_{\rm max}$ is a reasonable limit on the number of $e$-folds of inflation within the Gaussian approximation.
For the same reason, a bound derived by requiring that a non-negligible but smaller than ${\cal O}(1)$ fraction of patches are forming defects will not be significantly more constraining than ${\cal N}_{\rm max}$.\footnote{Likewise, while the divergence is assumedly unphysical and would be regulated if we cut off $\delta h$ by, \eg, throwing away backreacting patches, such a procedure would not affect these results until the fraction of backreacting patches became non-negligible at ${\cal N} \lsim {\cal N}_{\rm max}$.}

Second, as mentioned, we are working in the Hartree-Fock approximation, such that $\vev{\delta h^4} = 3 \vev{\delta h^2}^2$.  This holds if the operator $h$ is a Gaussian stochastic quantity, but breaks down for a field with self-interactions, $\lambda \neq 0$, as is the case for the Higgs field.  In particular, once the fluctuations become large (for $\delta h \gsim H$), the potential term in \Eref{eq:hsqevolution} and the self-interactions will become important.  For $\lambda < 0$, this will enhance the fluctuations, potentially increasing the proportion of unstable patches at any given $e$-fold---we return to this point in \Sref{sec:FP}.
In particular, we will see that self-interactions can drastically modify the behavior of the most unstable patches.  Consequently, while the limit ${\cal N}_{\rm max}$ is valid within the Gaussian approximation, an actual limit on ${\cal N}$ for the case of the SM Higgs may be substantially different, depending on the behavior of (and cosmological constraints on) the crunching patches.

Finally, we have considered $\lambda$ as constant and negative.  We argue why this choice is appropriate for the Higgs field with $H > \Lambda_I$ in the next section.

%%%%%%%%%%%%%%%%%%%%%%%%%%%%%%%%%
\section{Perturbative Calculation of $\vev{\delta h^2(t)}$ and Connection to Standard Model Higgs Boson}
\label{sec:2pt}
%%%%%%%%%%%%%%%%%%%%%%%%%%%%%%%%%

As shown in the previous section, scalar field fluctuations are initiated and grow due to the quantum noise from dS.  Moreover, for $V(h) = \frac{\lambda}{4} h^4$ with $\lambda < 0$, $\vev{\delta h^2(t)}$ diverges in finite time, signaling the breakdown of our slow-roll solution and an end of the usual inflationary scenario.
Although we have used the HF approximation for demonstration, the growth of the fluctuations can be captured by a perturbative calculation, which is consistent with the result \Eref{eq:HFSoln} within the
range that the perturbative calculation is valid.\footnote{The HF approximation effectively resums the leading IR logarithms that arise in the perturbative calculation of the two-point correlation function.  It has been shown that the stochastic approach of, \eg, \cite{Starobinsky:1994bd} gives IR logarithms consistent with a perturbative QFT calculation \cite{Woodard:2006pw,Tsamis:2005hd,Garbrecht:2013coa,Garbrecht:2014dca}.}  Perturbation theory eventually breaks down due to the logarithmic growth of scalar correlators---thus, by calculating $\vev{\delta h^2(t)}$ perturbatively, we can determine when the breakdown occurs and identify this with the singularity of \Eref{eq:HFSoln}, providing a non-trivial check of the results of the previous section.
Most importantly, the perturbative calculation will elucidate how to extend the results for the toy $\lambda h^4$ model considered in \Sref{sec:hartreefock} to the case of the full SM, our ultimate goal.

\begin{figure}
\includegraphics[width=0.9\textwidth]{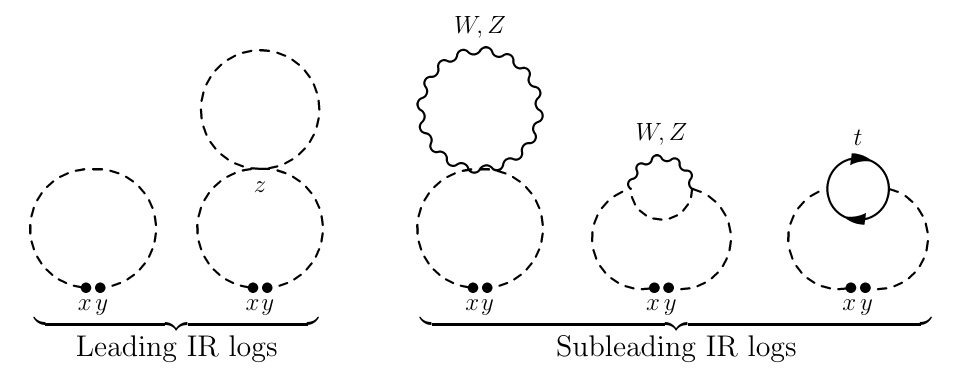}
\caption{Sample Feynman diagrams included in our calculation of the Higgs two-point correlation function.  The first two graphs (labeled ``Leading IR logs'') contribute to the late-time divergence of the Higgs two point correlation $\vev{\delta h^2(t)}$.  The last three graphs do not directly contribute to the leading divergence, but serve to renormalize the Higgs self-coupling $\lambda$. The points $x,~y$ are assumed to be separated by less than one Hubble length during inflation and the gauge boson propagators, for reasons we explain in the text, include only the transverse degrees of freedom.  \label{fig:FeynmanDiags}}
\end{figure}

We summarize in \Fref{fig:FeynmanDiags} the relevant diagrams for computing the evolution of the two-point correlation function in the SM.  We start by computing the first two graphs in \Fref{fig:FeynmanDiags} in $\lambda h^4$ theory. We will find that computing these first two graphs reproduces the leading behavior that we observed in the previous section.   We will also argue that the other graphs do not contribute to the leading divergence of the Higgs two-point correlation function.  This observation will allow us to connect our toy model to the SM.

To be explicit, we compute the two-point correlation function utilizing the ``in-in'' formalism. (For a review of the ``in-in'' formalism and its applications to cosmology, see \cite{Weinberg:2005vy}.)
The expectation value of an operator $\langle{\cal O}\rangle$ to a given order $n$ in perturbation theory is
\be
\left\langle \mathcal{O}(t)\right\rangle =\sum_{n}\left(-i\right)^{n}\int_{-\infty}^{t}dt_{1} \cdots \int_{-\infty}^{t_{n-1}}dt_{n} \left\langle \left[\left[\mathcal{O}^{I}(t),H^{I}(t_{n})\right],\cdots H^{I}(t_{1})\right]\right\rangle ,
\ee
where the superscript $I$ denotes that the operators are in the interaction picture, and $H^{I}$ is the interaction Hamiltonian density,
\begin{equation}
H^I = \frac{1}{4}\lambda\left(h^{I}(z)\right)^{4}+\frac{1}{2}\delta m^{2}\left(h^{I}(z)\right)^{2}+\frac{1}{2}\delta\xi R(z)\left(h^{I}(z)\right)^{2}.
\end{equation}
In this Hamiltonian we have included counterterms for the mass and curvature coupling, $\delta m^2$ and $\delta \xi$, both of which are induced through RG effects (see, for example, \cite{Herranen:2014cua}).  Our renormalization conditions will fix these quantities at $\mu = H$.

At next-to-leading order (\ie, including $n=0,~1$ contributions), we thus have
\begin{align}
\left\langle h(t,\vec{x})h(t,\vec{y})\right\rangle & = \left\langle h^{I}(t,\vec{x})h^{I}(t,\vec{y})\right\rangle \nonumber \\
& +\left(-i\right)\int_{-\infty}^{t}dt_{z}\sqrt{-g(t_{z})}\int d^{3}\vec{z}\left\langle \left[h^{I}(t,\vec{x})h^{I}(t,\vec{y}), \right. \right. \nonumber \\ & \qquad \left. \left. \frac{1}{4}\lambda\left(h^{I}(z)\right)^{4}+\frac{1}{2}\delta m^{2}\left(h^{I}(z)\right)^{2}+\frac{1}{2}\delta\xi R(z)\left(h^{I}(z)\right)^{2}\right]\right\rangle,
\end{align}
where $R$ is the Ricci scalar,
\be
R=12H^{2},
\ee
in dS spacetime.  The $n= 0$ contribution corresponds to the first graph in \Fref{fig:FeynmanDiags} and the $n=1$ contribution corresponds to the second graph in \Fref{fig:FeynmanDiags}.

Defining
\begin{align}
\rho(x,y) & = i\left\langle \left[h^{I}(x),h^{I}(y)\right]\right\rangle, & F(x,y) & = \frac{1}{2}\left\langle \left\{ h^{I}(x),h^{I}(y)\right\} \right\rangle ,
\end{align}
we have
\begin{align}
\left\langle h(t,\vec{x})h(t,\vec{y})\right\rangle & = F(x,y) - \int^{t}d^{4}z\, a^{3}(t_{z})\left[F(x,z)\rho(y,z)+\rho(x,z)F(y,z)\right] \left(3\lambda F(z,z)+\delta m^{2}+\delta\xi R(z)\right),\label{eq:2pt_function}
\end{align}
with
\be
F(x,y)=\frac{1}{2}\int\frac{d^{3}k}{\left(2\pi\right)^{3}}h_{k}(t_{x})h_{k}^{*}(t_{y})e^{i\vec{k}\cdot(\vec{x}-\vec{y})}+ \text{c.c.}
\ee

Let us start by calculating the scalar loop contribution to the correlator, encoded in the function $F(z,z)$ and shown in the second diagram of \Fref{fig:FeynmanDiags}.  Using the mode functions for a massless mode in dS space, \Eref{eq:modesolution}, we have
\begin{align}
3\lambda F(z,z) & = 3\lambda\int_{\Lambda_{IR}}^{a\Lambda}\frac{d^{3}k}{\left(2\pi\right)^{3}}\left|h_{k}(t_{z})\right|^{2} \label{eq:Fzz} \\
 & = 3\lambda\left[\frac{\Lambda^{2}}{8\pi^{2}}+\frac{H^{2}}{8\pi^{2}}\ln\left[\left(\frac{a\Lambda}{\Lambda_{IR}}\right)^{2}\right]\right] \label{eq:scalarloop},
\end{align}
where the IR cut-off is taken to be $\Lambda_{IR} = a_{0}H$, as in \Sref{sec:hartreefock}.  There are two types of terms present in \Eref{eq:scalarloop}.\footnote{We have dropped exponentially suppressed terms that go as $\Lambda_{\rm IR}/aH$.}  First, there are the IR logarithms of the form $\log(a/a_0) = {\cal N}$, due to the superhorizon modes, that give rise to the divergence of the correlator $\vev{\delta h^2}$ as observed in the previous section.  Second, there are terms due to UV physics, including quadratic and logarithmic divergences.  These terms are identical to terms that would be present in Minkowski space, as the high-energy subhorizon modes only see the local spacetime (which appears flat) and not the expansion.  As such, these terms can be cancelled by local counterterms $\delta m^2, \delta \xi$,
\begin{align}
\delta m^2 & = -3 \lambda(\mu) \frac{\Lambda^2}{8 \pi^2}, & 12 \delta \xi & = -\frac{3 \lambda(\mu)}{4 \pi^2}\log\left(\frac{\Lambda^2}{\mu^2}\right).
\end{align}
As in Minkowski space, the UV divergences are accompanied by logarithms of the renormalization scale and the energy scale $H$, $\log(\mu^2/H^2)$.
We have chosen a renormalization condition for the mass-squared and non-minimal coupling such that the renormalized $m^2(\mu)$ and $\xi(\mu)$ vanish at $\mu = H$.

Putting the pieces together, the correction to the two-point correlation function goes as
\be
3\lambda F(z,z)+\delta m^{2}+\delta\xi R=\frac{3\lambda(\mu) H^{2}}{8\pi^{2}}\left(2{\cal N}+\ln\frac{\mu^{2}}{H^{2}}\right).
\label{Eq:Fzz}
\ee
The choice of renormalization scale resums the logarithms and ensures the theory remains perturbatively under control in the UV---specifically, the logarithms vanish for the choice $\mu = H$, and the coupling is the RG-improved tree-level coupling $\lambda(\mu = H)$.
We note that the effects of the IR logarithms from higher-order corrections are also minimized by choosing $\mu = H$.
In the remainder of the calculation, we will be focused on extracting the leading IR logarithms, which determine the rate at which the two-point correlation diverges.
First, though, we note that this simple analysis suggests how contributions from additional Standard Model particles are to be included.  The contributions from loops of SM particles in the UV are shown as the ``subleading IR logs'' diagrams in \Fref{fig:FeynmanDiags}.   Loops of transverse gauge bosons and fermions actively renormalize the coupling $\lambda(\mu)$ from the UV cut-off of the theory $\Lambda$ down to $\mu = H$.  At scales below $\mu = H$, however, the propagators of these fields do not have logarithmic divergences at late time and hence do not contribute to the divergent part of $\vev{\delta h^2}$---we elaborate on this point further below.

The leading term in \Eref{eq:2pt_function} is
\begin{align}
F(t,\vec{x};t,\vec{y}) & = \frac{H^{2}}{4\pi^{2}}\left(\ln\frac{1}{\Lambda_{IR}r}+1-\gamma\right)+\frac{1}{2\pi^{2}}\frac{1}{a^{2}r^{2}},\\
 & \approx \frac{H^{2}}{4\pi^{2}}{\cal N}
\end{align}
with $r=\abs{\vec{x}-\vec{y}}$ evaluated at $r \approx (aH)^{-1}$, keeping the leading IR logarithm.
The leading IR logarithm due to second term of \Eref{eq:2pt_function} is
\begin{align}
& - 3\lambda\int^{t}d^{4}z\, a^{3}(t_{z})\frac{H^{2}}{8\pi^{2}}H\left(t_{z}-t_{0}\right)\left[F(x,z)\rho(y,z)+\rho(x,z)F(y,z)\right] \nonumber \\
& \qquad = \left(-i\right)3\lambda\int_{t_{z},k}a^{3}(t_{z})\frac{H^{2}}{8\pi^{2}}H\left(t_{z}-t_{0}\right)\left[u_{k}^{2}(t)u_{k}^{*2}(t_{z})e^{-i\vec{k}\cdot\left(\vec{x}-\vec{y}\right)}-\mbox{h.c.}\right] \nonumber \\
& \qquad \approx -\frac{\lambda}{24\pi^{2}}H^{2}{\cal N}^{3}.\label{eq:superhorizon_approx}
\end{align}
We can compare this with the result of \Eref{eq:HFSoln}, expanded in the limit of $\sqrt{-\lambda} {\cal N} \ll 1$,
\be
\vev{\delta h^2(t)}_{\rm HF} \approx \frac{H^2}{4 \pi^2} {\cal N} - \frac{\lambda H^2}{24 \pi^4} {\cal N}^3.
\label{eq:Nexpansion}
\ee
The two results agree, consistent with the claim that the HF approach resums the leading IR logarithms that arise in perturbation theory.\footnote{A similar analysis has been done in Refs.~\cite{Garbrecht:2013coa,Garbrecht:2014dca} using the stochastic approach.}

We see that perturbation theory breaks down (signaled by the subleading term exceeding the tree-level term) after a critical number of $e$-folds
\be
{\cal N} > \pi \sqrt{\frac{6}{\abs{\lambda}}} \equiv {\cal N}_c \gsim {\cal N}_{\rm max}.
\ee
Although we have only calculated the breakdown of perturbation theory at leading order, we can see that the result is consistent with ${\cal N}_{\rm max}$ derived from \Eref{eq:HFSoln}.  In addition, for $\lambda < 0$, the subleading term gives a positive contribution to $\vev{\delta h^2(t)}$, further supporting the claim that the correlator diverges in finite time.\footnote{The perturbative calculation also breaks down in finite time for $\lambda > 0$.  This corresponds to the fluctuations approaching a stabilized, equilibrium solution \cite{Starobinsky:1994bd}---this solution is also apparent in the late-time limit of \Eref{eq:HFSoln} with $\lambda > 0$.}

In this analysis, the leading IR logarithmic divergences play the crucial role. They originate from (1) the time integral associated with the scalar correlation functions $F$ and $\rho$ in the superhorizon limit, and (2) the spatial momentum integral of the superhorizon modes involving the correlation function $F$.
In other words, as the superhorizon modes of a minimally-coupled light scalar are undamped, its correlation functions are enhanced by scalar loop corrections, leading to logarithmic growth with the scale factor $a$.\footnote{Note that we obtained a higher power of logarithmic divergence from $n=1$ than from $n= 0$, suggesting that perturbatively higher order diagrams involving more scalar propagators (and thus more time and momentum integrals) generally give a rise to higher powers of the IR divergence.}
By contrast, fermions and transverse gauge bosons have decaying superhorizon mode functions (in the IR).  Woodard and collaborators (see, for example, \cite{Tsamis:2005hd,Prokopec:2006ue,Woodard:2006pw,Prokopec:2007ak} and references therein) accordingly classified minimally-coupled light scalar fields as {\em active} fields and others as {\em passive}---diagrams involving the passive fields, such as the last three graphs in \Fref{fig:FeynmanDiags}, do not contribute to the leading IR divergence.  As a result, diagrams contributing to the leading IR divergence at a given order in perturbation theory are composed solely of scalar propagators.  This is the first key observation that allows us to connect our toy model to the SM---if we are only interested in extracting the leading IR divergence, the calculation including only the scalar field still applies in the exactly same way.

The second observation crucial to connect our calculation to the SM is that, while the additional SM fields do not contribute to the leading IR logs directly, they do renormalize the quartic coupling in the ultraviolet (UV), and hence determine $\lambda(\mu)$.
In particular, as hinted at in the calculation and discussion below \Eref{Eq:Fzz}, $\lambda$ for the SM should be chosen as {\em the RG-improved SM quartic evaluated at $\mu = H$, $\lambda(H)$}.  The basic idea is that, since SM fermions and vector bosons cannot generate the leading IR logarithms, when including all of the diagrams as in \Fref{fig:FeynmanDiags} one would obtain schematically
\begin{align}
\vev{\delta h^2(t)}& = \frac{\hbar H^2}{4 \pi^2} \left( {\cal N} + c_1 \right) - \frac{\hbar^2\lambda H^2}{24 \pi^4}  \left( {\cal N}^3 + d_1  {\cal N}^2 + d_2 {\cal N} + d_3 \right) + \mathcal{O}(\hbar^3),
\end{align}
where we have restored $\hbar$ to show the order in perturbation theory.  The contributions from additional SM particles are encoded in the coefficients of the subleading IR logs, $c_1$ and $d_i$.  These contributions are minimized by choosing an optimal renormalization scale $\mu$.  In particular, the toy model analysis and the effective potential in curved spacetime \cite{Herranen:2014cua} indicates the optimal renormalization scale is $\mu \approx H$.  Said another way, the curvature $R=12H^2$ effectively plays the role of the UV cut-off of the scalar-only theory.

So, in summary, the dynamics of the superhorizon Higgs fluctuations are captured by both mode evolution in the Hartree-Fock approximation and perturbative calculation for a simple scalar model with $V(h) = \frac{\lambda}{4} h^4$, where $\lambda = \lambda(\mu = H)$ is the RG-improved SM quartic evaluated at $\mu = H$.
To complete the connection, some comments are in order regarding the additional scalar degrees of freedom of the SM Higgs multiplet and the gauge invariance of this analysis.  One may explicitly calculate the gauge-invariant (composite) operator $\vev{{\cal H}^\dagger {\cal H}}$, where ${\cal H}$ is the full SM Higgs multiplet,
\be
{\cal H} = \frac{1}{\sqrt{2}} \left(\begin{array}{c} \chi_1 + i \chi_2 \\ \bar{h} + \delta h + i \chi_3 \end{array}\right).
\ee
Notably, ${\cal H}$ contains additional light bosonic degrees of freedom that should experience a similar growth in fluctuations and contribute to correlators (analogous to the contributions of the would-be GBs in SQED observed in \cite{Prokopec:2006ue,Woodard:2006pw}).  While these degrees of freedom are eaten by the SM gauge bosons in a background with $\vev{{\cal H}^\dagger {\cal H}} \neq 0$, like the Higgs they remain effectively massless as long as $g^2 \vev{\delta h^2(t)} \lsim H^2$ (where $g$ represents the SM gauge coupling), and so should exhibit $\vev{\delta \chi_i^2(t)} \approx \vev{\delta h^2(t)}$, at least during the early stages of inflation.  Accounting for these contributions in the Hartree-Fock approximation, the mode equation for the superhorizon modes becomes (taking $V({\cal H}) = \lambda ({\cal H}^\dagger {\cal H})^2$)
\be
3 H \dot{h}_k(t) + \lambda \left(3 \vev{\delta h^2(t)} + \sum_i \vev{\delta \chi_i^2(t)}\right) h_k(t) \approx 3 H \dot{h}_k(t) + 6 \lambda \vev{\delta h^2(t)} h_k(t) = 0.
\ee
Comparing with \Eref{eq:shmodeeom}, we see that, if the $\chi_i$ fields remained light, $\vev{\delta h^2(t)}$ would effectively diverge as if it had a factor of $2$ larger coupling,
\begin{align}
\label{eq:HFwithGBs}
\vev{{\cal H}^\dagger {\cal H}} & = \frac{1}{\sqrt{-\lambda}} \frac{H^2}{2\pi} \tan \left(\sqrt{-\lambda} \frac{{\cal N}}{\pi}\right), & {\cal N}_{\rm max} & = \frac{\pi^2}{2 \sqrt{-\lambda}}.
\end{align}
Similarly, including the additional degrees of freedom (for instance, in the second diagram of \Fref{fig:FeynmanDiags}) in the perturbative calculation we have carried out in this section, \Eref{eq:Nexpansion} becomes
\be
\label{eq:PTwithGBs}
\vev{{\cal H}^\dagger {\cal H}} \approx \frac{H^2}{2 \pi^2} {\cal N} - \frac{\lambda H^2}{6 \pi^4} {\cal N}^3.
\ee
Again, the two results agree in the limit $\sqrt{-\lambda} {\cal N} \ll 1$.  Note that, although the $\chi_i$ accelerate the divergence initially, their effects should decouple as $g^2 \vev{{\cal H}^\dagger {\cal H}}$ becomes comparable to $H^2$.  Moreover, as $g^2 \gg \lambda$, this decoupling occurs before the breakdown of the perturbative expansion---in particular, terms of $\mathcal{O}(\hbar^3 \lambda g^2)$ will become relevant in \Eref{eq:PTwithGBs}, canceling against the subleading term.  Consequently, for the SM, the appropriate limit  ${\cal N}_{\rm max}$ in the Gaussian approximation should lie somewhere between \Eref{eq:Nmax} and \Eref{eq:HFwithGBs}.

Having made the connection between the simplified model of \Sref{sec:hartreefock} and the full Standard Model via the perturbative calculation, we finish this section by commenting on the phenomenological implications of the upper bound on the number of $e$-folds, ${\cal N} \leq {\cal N}_{\rm max}$.
As argued, the appropriate numerical value for $\lambda$ for the case of the SM Higgs is $\lambda(\mu = H)$.  At scales $\mu \gg \Lambda_I$, the quartic coupling in the SM approaches a conformal regime with $\lambda \approx - 0.01$ for the best-fit values of the relevant parameters.  The corresponding limit on inflation is
\be
50 \lsim {\cal N}_{\rm max} \lsim 70,
\ee
where the lower (upper) limit corresponds to treating the $\chi_i$ as always light (decoupled).\footnote{Within the HF approximation, one can estimate the impact of realistic $\chi_i$ decoupling by treating $\chi_i$ as light until $g^2 \vev{\delta h^2(t)} \approx H^2$---doing so, one finds that the divergence is delayed by $\sim$ 15-20\% relative to \Eref{eq:HFwithGBs}.}
This is intriguingly close to the ${\cal N}_o \sim$ 50-60 $e$-folds required for consistency with cosmic microwave background (CMB) observations.  Thus, if the Hubble scale during inflation is much larger than the instability scale, $H \gg \Lambda_I$, new physics may be required to stabilize the Higgs potential and make our Universe observationally viable.  The imminent discovery of primordial $B$-modes in the CMB would therefore merit a more precise determination of ${\cal N}_{\rm max}$.  For $\lambda \approx -5 \times 10^{-3}$, as perhaps appropriate for $H \gsim \Lambda_I$,
\be
70 \lsim {\cal N}_{\rm max} \lsim 100,
\ee
such that our Universe could perhaps arise after 50-60 $e$-folds of inflation even if the Higgs potential is unstable.  Note again, just as we did for the HF approximation in \Sref{sec:hartreefock}, that the proper limit for the SM Higgs may depend on the dynamics of the most unstable patches in which non-Gaussian nature of the field is relevant---we address this issue in \Sref{sec:FP}.

As observed at the end of \Sref{sec:hartreefock}, inflating fewer than ${\cal N}_{\rm max}$ $e$-folds is a necessary, but not sufficient, condition for a successful inflationary epoch.  If $\vev{\delta h^2(t_e)} \gsim \Lambda_I^2$, where $t_e$ is the time at the end of inflation, a majority of patches have Higgs vevs in the unstable regime at the end of inflation.  Patches with $\delta h > (<) \; \Lambda_{\rm max}$ subsequently evolve towards the true (electroweak) vacuum and, as the horizon expands post-inflation, the different patches come back into causal contact with one another.  This gives rise to a Universe with regions of different Higgs vev separated by domain walls, in which the lower-energy-density true vacuum regions would percolate and come to dominate space, again precluding a Universe such as ours.
Indeed, the existence of a single true vacuum patch at the end of inflation may be sufficient to overwhelm the electroweak patches, making our Universe unlikely even if such patches are extremely rare as a result of the huge number of patches $e^{3 \mathcal{N_{\rm end}}}$ present at the end of inflation \cite{Hook:2014uia,Fairbairn:2014zia}.  However, we avoid this situation by having a sufficiently high re-heat temperature, $T_R^2 \gtrsim  \vev{\delta h^2(t)}$.  The Higgs then becomes rapidly thermalized and settles down to the electroweak vacuum.

We have now shown how to compute the upper bound on the number of $e$-folds that inflation can proceed before large local field fluctuations produce large inhomogeneities, precluding a relatively homogeneous Universe such as ours.  So far we have only done this either assuming a Gaussian distributed field (\Sref{sec:hartreefock}), or carrying out a perturbative expansion that breaks down just as the instabilities become important (this section).  In the next section, we consider the Fokker-Planck equation that, once supplied with the correct potential, reproduces the non-Gaussian tails of the distribution and allows us to gain more information about the rare but important unstable patches.  This will in turn allow us to better understand the Universe that emerges.

%%%%%%%%%%%%%%%%%%%%%%%%%%%%%%%%%%%%%%%%
\section{Standard Model Higgs in the Fokker-Planck Equation}
\label{sec:FP}
%%%%%%%%%%%%%%%%%%%%%%%%%%%%%%%%%%%%%%%%%

The Fokker-Planck (FP) approach to studying the evolution of scalar field fluctuations in a dS background was previously applied to the Higgs in \Refs{Hook:2014uia,Fairbairn:2014zia,Espinosa:2007qp}.  Here we make use of what we learned in \Srefs{sec:hartreefock}{sec:2pt} about Higgs potential during inflation to make contact with previous results, notably those in \cite{Hook:2014uia}.  We will not find significant numerical differences with \Ref{Hook:2014uia}, but we will be able to better interpret those results.

The FP equation,
\be
\label{eq:FPeqn}
\frac{\partial P}{\partial t} = \frac{\partial}{\partial \delta h}\left[\frac{V'(\delta h)}{3 H} P + \frac{H^3}{8 \pi^2} \frac{\partial P}{\partial \delta h}\right],
\ee
describes the evolution of a probability distribution function, $P(\delta h,t)$, which can be interpreted as the probability for the field to take a value $\delta h$ in a Hubble patch at time $t$.  The first term on the right-hand side is a drift term due to the external potential, while the second term is a diffusion piece due to quantum fluctuations of the Higgs in an inflationary background.
$P(\delta h,t)$ can then be used to calculate superhorizon correlation functions via
\be
\vev{\delta h^n(t)} = \int d\delta h \, (\delta h)^n P(\delta h,t).
\ee
This formalism is intended to capture the non-trivial infrared behavior exhibited in dS space by scalar field fluctuations and correlators such as $\vev{\delta h^2(t)}$ (as considered in \Sref{sec:hartreefock}) \cite{Starobinsky:1994bd}.

As stated in the introduction, one important question for the SM Higgs boson is which potential $V(h)$ to use.
From the perturbative approach of \Sref{sec:2pt}, it is clear that the leading divergent behavior of field distributions and correlators is captured by a stochastic description of field dynamics (such as HF) if one simply uses a tree-level quartic potential with constant coupling, taken to be the RG-improved quartic coupling evaluated at the scale $H$.
This resums UV logarithms of the form $\log(\mu^2/H^2)$ that appear in perturbation theory and as such, in the case of the Higgs, encodes the local, subhorizon effects of the SM gauge bosons and fermions, which decouple on superhorizon scales.
Consequently, the results of the FP equation solved for a model with $V(h) = \frac{\lambda^\prime}{4} h^4$ and the identification $\lambda^\prime \approx \lambda(H)$ should describe well the dynamics of the Higgs field fluctuations during inflation.  In particular, this prescription should unambiguously capture the leading divergent behavior.

\begin{figure}
\includegraphics[width=\textwidth]{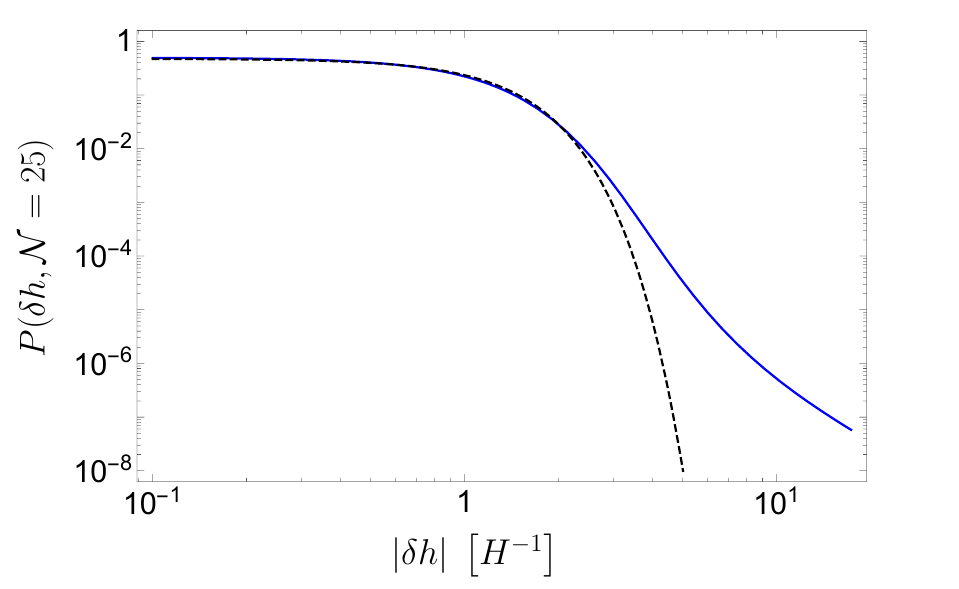}
\vspace{-0.05\textheight}
\caption{\label{fig:Pplot} Probability distribution of the Higgs, $P(\delta h,t)$, evaluated at ${\cal N} = 25$ for the case of $\lambda(H) = -0.01$ (blue, solid).  Also shown is a Gaussian distribution, corresponding to the Hartree-Fock approximation of \Sref{sec:hartreefock}, with variance $\vev{\delta h^2(t)}$ given by \Eref{eq:HFSoln} (black, dashed).}
\end{figure}

The advantage of the FP approach relative to the HF approximation employed in \Sref{sec:hartreefock} is that the FP equation incorporates non-Gaussianity, which is relevant for any self-interacting scalar field, particularly at large field values.
In the case of an unstable potential, such as that of the Higgs with $H > \Lambda_I$, self-interactions can accelerate the rate at which the fluctuations diverge in a patch, producing long tails in the distribution $P(\delta h, t)$.
These tails can be seen in \Fref{fig:Pplot}, which shows $P(\delta h, t)$ after ${\cal N} = 25$ $e$-folds of inflation for $\lambda(H) = -0.01$.
At small $\abs{\delta h}$, the dynamics are dominated by the stochastic noise term, and the distribution broadens steadily over the course of inflation, as in the Gaussian approximation.  However, at large $\abs{\delta h}$---specifically, for
\be
\delta h \gsim \delta h_{\rm classical} \equiv \left(\frac{3}{-2 \pi \lambda}\right)^{1/3} H
\ee
---the classical force due to the potential, $V^\prime(h) = \lambda h^3$ comes to dominate over the quantum fluctuations, causing the tails of the distribution to spread out rapidly.  For comparison, we also show a Gaussian distribution with variance $\vev{\delta h^2(t)}$ given by \Eref{eq:HFSoln}; the distributions are similar for $\abs{\delta h} \lsim \delta h_{\rm classical} \approx 4 H$, but the FP distribution exhibits higher probability to find the field at larger values $\abs{\delta h} \gsim \delta h_{\rm classical}$.

Note that one potential subtlety does arise in solving the FP equation for the SM Higgs due to the presence of the additional $\chi_i$ bosonic degrees of freedom in the full Higgs multiplet, discussed at the end of \Sref{sec:2pt}---specifically, they potentially obfuscate the most appropriate choice of $\lambda^\prime$ for best replicating Higgs behavior during inflation.
For instance, following \Erefs{eq:HFwithGBs}{eq:PTwithGBs}, $\lambda^\prime \approx 2 \lambda(H)$ would be the correct choice if the $\chi_i$ remain lighter than $H$.  But, as their mass $m_{\chi_i}^2 \sim g^2 \vev{\delta h^2}$ becomes important, their effects decouple, so that this is not the correct prescription at late time.  In fact, because the impact of the potential only becomes significant once $\abs{\delta h} \gsim \delta h_{\rm classical}$, at which point $m_\chi \sim g \, \abs{\delta h} \gsim H$, the choice $\lambda^\prime = \lambda(H)$ should be a better approximation.  Thus, we concentrate on this choice here.

The regions of the distribution $P(\delta h, t)$ with large $\delta h$ contribute significantly to correlation functions, causing them to diverge rapidly---much more rapidly than in the Gaussian approximation, for instance.  While one would reasonably expect Higgs self-interactions to accelerate the rate of divergence, there is a question as to what extent this divergence is physical.  The evolution of $\delta h$ to arbitrarily large value is clearly unphysical---at very least, for $\abs{\delta h} \sim M_P$, Planck-suppressed operators would influence the evolution of $\delta h$.  Moreover, as mentioned in \Sref{sec:hartreefock}, once $\delta h$ exits the slow-roll regime in a patch when
\be
\abs{\delta h} \gsim \delta h_c \equiv \left(\frac{3}{-\lambda}\right)^{1/2} H,
\ee
the slow-roll approximation employed by the FP equation breaks down, the vev in the patch quickly diverges, and the patch evolves to a singularity.
Such patches effectively disappear when they crunch, so it is not clear that they should be included in $P(\delta h, t)$ or when calculating $\vev{\delta h^n(t)}$.  However, truncating the probability distribution at a particular value of $\delta h$ will of course cut off the divergence of the correlators, in contrast to the Hartree-Fock approach where backreaction and the disappearance of patches is neglected.
Consequently, determining when our Universe stops being viable based on the divergence of, \eg, $\vev{\delta h^2(t)}$ or $\vev{V(\delta h)}$ as in \Sref{sec:hartreefock} is no longer sensible, and we instead need an alternative prescription to interpret $P(\delta h, t)$ for the fate of our Universe.

One reasonable approach, as employed in \Refs{Espinosa:2007qp,Hook:2014uia}, is to assume that any patches that transition to the unstable regime benignly crunch during inflation, and thus to concentrate on the proportion of patches $P_\Lambda$ that survive in the electroweak vacuum at the end of inflation (see \Eref{eq:survivalprobability}).
In order to do so, based on the observation that $\delta h$ diverges rapidly and $P(\delta h,t)$ flattens out to small values for $\abs{\delta h} \gsim \delta h_{\rm classical}$, one can solve the FP equation approximating $P(\abs{\delta h} \ge \Lambda_c,t) = 0$, where $\Lambda_c \geq \delta h_{\rm classical}$, as in \Ref{Hook:2014uia}.  This prescription well captures the bulk of the distribution at $\abs{\delta h} < \delta h_{\rm classical}$ and so is suitable for calculating $P_\Lambda$.
Such an analysis reveals that inflation can always last long enough to replace the lost patches.
So, in this case, the instability in the SM Higgs potential does not abort inflation or preclude the existence of our Universe.

However, as discussed previously, if a significant proportion of crunching or unstable patches are present at the end of inflation, the resulting universe might not look like \emph{our} Universe.  While unstable patches with $\abs{\delta h} > \Lambda_{\rm max}$ can be stabilized by efficient reheating (see the discussion at the end of \Sref{sec:hartreefock}), the defects and large inhomogeneities formed by crunching patches with $\abs{\delta h} \gsim \delta h_c$ {\em at the end of inflation} may well be inconsistent with the small curvature perturbations observed in our Universe.
Moreover, if the proportion of crunching patches becomes ${\cal O}(1)$, inflation is expected to terminate altogether as space becomes dominated by collapsing regions \cite{Sekino:2010vc}.
Consequently, in this work, we employ a slightly different approach to \Ref{Hook:2014uia}, and concentrate on the minimal level of inhomogeneity one would expect to be generated at any point during inflation due to the Higgs instability.
Specifically, we numerically solve the FP equation to determine the proportion of surviving patches that are transitioning out of the slow-roll regime at each $e$-fold of inflation,
\be
\label{eq:fNdef}
f_{{\cal N}} \equiv \frac{\int_{-\delta h_c}^{\delta h_c} d \delta h \left\{P(\delta h, {\cal N}) - P(\delta h, {\cal N} - 1)\right\}}{\int_{-\delta h_c}^{\delta h_c} d \delta h P(\delta h, {\cal N} - 1)}.
\ee
The extremely rapid crunching of these patches would likely give rise to defects at the end of inflation even if efficient reheating occurred.

\begin{figure}
\includegraphics[width=\textwidth]{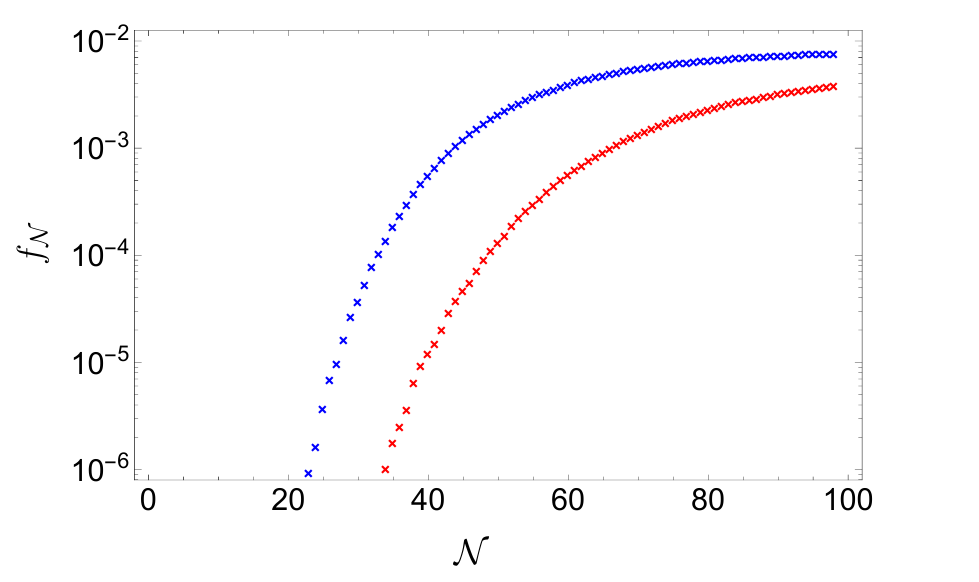}
\vspace{-0.05\textheight}
\caption{\label{fig:fNplot} The proportion of surviving patches transitioning out of the slow-roll regime $f_{\cal N}$ (\Eref{eq:fNdef}) at ${\cal N}$ $e$-folds of inflation for two choices of quartic coupling $\lambda(H) = -0.005$ (red) and $-0.01$ (blue).}
\end{figure}

In \Fref{fig:fNplot}, we show $f_{{\cal N}}$ as a function of ${\cal N}$ for two different choices of $\lambda(H)$.
After a certain number of $e$-folds, the proportion of patches transitioning out of the slow-roll regime begins to drastically increase, before eventually asymptoting to a steady state where approximately the same proportion of patches form defects at any given ${\cal N}$.
The number of $e$-folds, ${\cal N}_{\rm FP}$, at which $f_{\cal N}$ first reaches a particular value is given in \Tref{tab:fNtab} for several choices of $\lambda(H)$.
For comparison, we show the limit on the number of $e$-folds, ${\cal N}_{\rm max, HF}$, that one determines from considering the divergence $\vev{\delta h^2(t)}$ in the Hartree-Fock approximation (see \Sref{sec:hartreefock}).
The lower (upper) bound on ${\cal N}_{\rm max}$ corresponds to treating the additional $\chi_i$ degrees of freedom in the full SM Higgs multiplet as light (heavy and decoupled) throughout inflation---a limit derived incorporating realistic decoupling of $\chi_i$ likely lies $\sim$ 15-20\% above the lower bound.
Finally, we show the number of $e$-folds at which a particular $f_{\cal N}$ is reached for a Gaussian distribution with variance given by \Eref{eq:HFSoln}, ${\cal N}_{\rm HF}$, in order to explicitly demonstrate the claim in \Sref{sec:hartreefock} that, in the Hartree-Fock approximation, a negligible proportion of patches are forming defects until ${\cal N}$ approaches ${\cal N}_{\rm max}$.

\begin{table}
\renewcommand{\arraystretch}{1}
\setlength{\tabcolsep}{12pt}
\begin{tabular}{ c || c || c c | c c | c c }
\multirow{2}{*}{$\lambda(H)$} & \multirow{2}{*}{${\cal N}_{\rm max, HF}$} & \multicolumn{2}{|c|}{$f_{{\cal N}} = 10^{-5}$} & \multicolumn{2}{|c|}{$f_{{\cal N}} = 10^{-3}$} & \multicolumn{2}{|c}{$f_{{\cal N}} = 10^{-2}$} \\
& & ${\cal N}_{\rm FP}$ & ${\cal N}_{\rm HF}$ & ${\cal N}_{\rm FP}$ & ${\cal N}_{\rm HF}$ & ${\cal N}_{\rm FP}$ & ${\cal N}_{\rm HF}$ \\ \hline
-0.005 & $70 \lsim {\cal N}_{\rm max} \lsim 99$ & 40 & 95 & 60 & 96 & -- & 97 \\
-0.010 & $49 \lsim {\cal N}_{\rm max} \lsim 70$ & 27 & 66 & 44 & 67 & -- & 68 \\
-0.015 & $40 \lsim {\cal N}_{\rm max} \lsim 57$ & 22 & 53 & 35 & 55 & 86 & 55
\end{tabular}
\caption{\label{tab:fNtab} Number of $e$-folds at which $f_{\cal N}$ first reaches a specific value ($10^{-5},~10^{-3},~10^{-2}$), as computed by the FP equation, ${\cal N}_{\rm FP}$, or by a Gaussian distribution with variance given by \Eref{eq:HFSoln}, ${\cal N}_{\rm HF}$.  Dashes denote that the asymptotic value of $f_{\cal N}$ is smaller than that given.  Also shown is the range of ${\cal N}_{\rm max,HF}$ derived from the divergence of $\vev{\delta h^2(t)}$ in the HF approach.  Lower (upper) bounds correspond to light (heavy and decoupled) $\chi_i$ as in \Eref{eq:Nmax} (\Eref{eq:HFwithGBs}).  To ascertain when $f_{\cal N}$ reaches a particular value relative to when $\vev{\delta h^2(t)}$ diverges, ${\cal N}_{\rm HF}$ should be compared to the upper limit.  See text for more details.}
\end{table}

As previously discussed, a legitimate assumption is that, in order to avoid large inhomogeneities and thus produce our Universe, inflation must end before a significant proportion of patches are forming defects during each $e$-fold.  In the case of the FP approach, this translates into the requirement that $f_{\cal N}$ at the end of inflation be smaller than some critical value, $f_{\cal N}^{\rm crit}$, which constrains the duration of inflation, ${\cal N} \leq {\cal N}^{\rm crit}$.
Conceptually, this is equivalent to requiring ${\cal N} \leq {\cal N}_{\rm max}$ in the Hartree-Fock approach of \Sref{sec:hartreefock}---as \Tref{tab:fNtab} demonstrates, requiring $f_{\cal N} \leq f_{\cal N}^{\rm crit}$ in the Gaussian approximation would yield ${\cal N}^{\rm crit} \approx {\cal N}_{\rm max}$ for most reasonable values of $f_{\cal N}^{\rm crit}$.

In terms of such a constraint, the main difference between the HF and FP approaches is that the non-Gaussian self-interactions captured by the FP approach cause $f_{\cal N}$ to reach non-negligible values much more rapidly, and well before the bulk of the distribution has spread out significantly in the Gaussian approximation.  Consequently, if the formation of a small proportion of defects at the end of inflation is prohibited (\ie, if $f_{\cal N}^{\rm crit}$ is small), the FP approach indicates that the existence of an instability in the Higgs potential is very likely inconsistent with our observed Universe.
However, another notable difference between the HF and FP limits is that, if $f_{\cal N}$ from the FP calculation asymptotes to a value $f_{\infty} < f_{\cal N}^{\rm crit}$, then the instability in the SM Higgs potential does not appear to preclude our Universe---a longer period of inflation will be sufficient to replace the crunching patches and dilute away the defects, consistent with the results of \cite{Hook:2014uia}.
Notably, as $f_{\infty} \lsim 10^{-2} \ll {\cal O}(1)$ for the representative values of $\lambda(H)$ considered, the Higgs instability does not appear capable of aborting inflation.

Of course, the key question is {\em what proportion of patches can be forming defects at the end of inflation such that the resulting inhomogeneities are still consistent with the observed Universe?}  Or, in other words, what is an appropriate value for $f_{\cal N}^{\rm crit}$?
One well-motivated guess is that the defects produced are Primordial Black Holes (PBHs), and a variety of limits on PBHs have been determined for different ranges of masses and lifetimes (see, \eg, \cite{Carr:2005zd,Green:2014faa} and references therein).  However, for relatively light PBHs ($M_{\rm BH} \lsim 10^6 \text{ g} \approx 6 \times 10^{29} \text{ GeV}$) that are formed early and evaporate quickly, as we expect to be the case for those generated by the crunching patches, potential constraints are limited.  If one assumes that evaporating PBHs leave Planck-mass relics, then one can obtain a bound by requiring that these relics do not overclose the Universe \cite{MacGibbon:1987my}.  In this case, the resulting constraint on the fractional energy density that can be contained in PBHs at the time of their formation and evaporation is very stringent because the relics dilute like matter and so their relative abundance increases during radiation-domination.
Since we expect the energy density in crunching patches to be comparable to that in surviving patches, this would likely imply a very small value of $f_{\cal N}^{\rm crit} \ll 10^{-10}$ for best-fit values of the relevant parameters.\footnote{This estimate assumes $H \geq \Lambda_I \gsim 10^{11} \text{ GeV}$ and $T_R \gsim H$ (such that unstable patches are stabilized during reheating).  Intriguingly, such a bound favors lower values of $T_R$ (and hence $\Lambda_I$), which would give a shorter period of the relative abundance of relics to increase.}
If PBH evaporation does not produce relics, then the radiation from the PBHs simply contributes to the radiation at reheating, and the resulting universe is consistent with ours provided reheating stabilizes any remaining unstable (but uncrunched) patches.

We thus return full circle to \Ref{Hook:2014uia}.  We started our discussion here by considering the toy model of \Sref{sec:hartreefock}, which illuminated how to think about infrared divergences in scalar field correlation functions.  This calculation reproduced the Gaussian bulk of the scalar field vev probability distribution over the $e^{3 \cal N}$ Hubble volumes produced during ${\cal N}$ $e$-folds of inflation, but left unclear how to connect to the SM Higgs.  We then turned to the perturbative calculation of \Sref{sec:2pt}, which showed how to connect the toy scalar model to the SM Higgs, though the calculation was limited to two loops.  The results of these two calculations did, however, indicate how to correctly apply the FP equation to the SM Higgs case. Unlike the other approaches, the FP equation computes the evolution of the non-Gaussian tails and re-sums the contributions from higher loops in perturbation theory.  Numerically we obtain similar results to \cite{Hook:2014uia}, but we have gained insight on the proper application of the formalism and results.

%%%%%%%%%%%%%%%%%%%%%%%%%%%%%%%%
\section{Conclusions}
\label{sec:conclusions}
%%%%%%%%%%%%%%%%%%%%%%%%%%%%%%%%

We have examined the evolution of the Higgs field during inflation in the case of an unstable Higgs potential (\ie, the quartic coupling runs negative) and a Hubble parameter during inflation $H$ that is larger than the scale $\Lambda_I$ at which the potential becomes unstable.  We applied new methods that both allowed us to systematically deal with the gauge dependence in the SM Higgs potential, and to understand how to apply our results to the evolution of the spacetime as a whole.
In particular, we found that the leading IR divergent behavior of Higgs fluctuations is captured by the Fokker-Planck equation solved for the tree-level potential $V(h) = \frac{\lambda}{4} h^4$, where $\lambda$ is the RG-improved Standard Model quartic evaluated at a scale $\mu = H$.
As in our previous work \cite{Hook:2014uia}, we found that the instability in the Higgs potential does not terminate inflation, even when $H \gg \Lambda_I$.

However, we do find that, as inflation proceeds, a larger and larger fraction of the patches develop an instability and even crunch in each $e$-fold of inflation.
For typical values of the SM Higgs quartic coupling, approximately $10^{-3}$ to $10^{-2}$ of the patches would be destroyed during the last $e$-fold of inflation.
The defects produced by these crunching patches could yield large inhomogeneities such that the resulting Universe would not look like ours.
Moreover, inasmuch as inflation usually dilutes away any unwanted defects, the Higgs instability can regenerate defects at the end of inflation.

The exact level of Higgs-instability-related defects that can be tolerated depends very much on the nature of these defects.  For instance, some unstable patches are expected to crunch and yield light Primordial Black Holes.  If the rapid evaporation of these Primordial Black Holes leaves Planck-mass relics, there are very stringent constraints from requiring the relics not exceed the present energy density in the Universe.  On the other hand, if no relics remain from evaporation, the Primordial Black Hole evaporation simply contributes to the radiation during reheating, and the resulting universe may indeed look like ours.  Thus, our conclusion is that the Higgs instability need not be fatal to high scale inflation.  We reserve a closer examination of the post-inflationary evolution for future work.

%\newpage

{\em Acknowledgments:} We thank Anson Hook, Bibhushan Shakya and Moira Gresham for collaboration at the initial stages of this project.  It is also our pleasure to thank Nima Arkani-Hamed, Tim Cohen, Daniel Chung, Michele Papucci, David Pinner, and Matt Schwartz for useful conversations. HY and KZ are supported by the DoE under contract DE-AC02-05CH11231.
JK is supported by the DoE under contract number DE-SC0007859 and Fermilab, operated by Fermi Research Alliance, LLC under contract number DE-AC02-07CH11359 with the United States Department of Energy.

%%%%%%%%%%%%%%%%%%%%%%%%%%%%%%%%%%%
\bibliography{Higgs_Inflation_Bibliography}

\end{document}